# Non static cosmic strings in Lyra geometry

**FAROOK RAHAMAN**

**Khodar Bazar , Baruipur – 743302,
24-Parganas (South), West Bengal,
India.
E-mail : farook_rahaman@yahoo.com**

Abstract :

The gravitational field of both local and global non static cosmic strings in the context of Lyra geometry are investigated. Local strings are characterized by having an energy momentum tensor whose only non null components are $T_t^t = T_z^z$ . As linearized Einstein equations are formally analogous to the Maxwell equations, the exterior solution does not depend on the radial distribution of the source and hence a Dirac δ function was used to approximate the radial distribution of the energy momentum tensor for a local cosmic string along the z-axis: $T_a^b = \delta(x) \delta(y) \text{diag} (\sigma, 0, 0, \sigma)$ , σ being the energy density of the string [A.Vilenkin. Phys.Rep.(1985)121,263]. For a global string, the energy momentum tensor components are calculated from the action density for a complex scalar field ψ along with a Maxican hat potential. The gravitational field of the global string is shown to be attractive in nature.

## Introduction:

The origin of structure in the Universe is one of the greatest cosmological mysteries even today. Cosmologists are generally assumed that at very early stages of its evolution , the Universe has gone through a number of phase transitions. One of the immediate consequences of this phase transitions is the formation of defects or mismatches in the orientation of the Higgs field in causally disconnected regions [1].
Among the topological defects cosmic strings have received particular attention mainly because of their cosmological implications [2]. The double quasar problem can well be explained by strings and galaxy formation might also be generated by density fluctuation in the early universe due to strings [2]. Depending on whether the symmetry that is broken during the phase transition is local or global, the corresponding topological defects are called local or global strings. At first, Vilenkin [3] has calculated the metric around a local string in the linear approximation of general relativity. Local strings are characterized by having energy momentum tensor components $T_r^r = T_\theta^\theta = 0$ and $T_t^t = T_z^z \neq 0$ while the global strings have non zero energy momentum tensors throughout entire space due to the presence of a goldstone boson field extending beyond the core.
In general theory of relativity , there have been a large amount of discussions on the gravitational field of static [4] and non static [5] strings beginning with the work of Vilenkin [3] .

In last few decades , there has been considerable interest in Alternative theories of gravitation .The most important among them are scalar-tensor theories proposed by Lyra [6] and by Brans-Dicke[6]. Lyra [6] proposed a modification Riemannian geometry by introducing a gauge function in to the structure less manifold that bears a close resemblance to Weyl's geometry . In general relativity Einstein succeeded in geometrising gravitation by identifying the metric tensor with the gravitational potentials . In the scalar-tensor theory of Brans-Dicke , on the other hand , scalar field remains alien to the geometry . Lyra's geometry is more in keeping with the spirit of Einstein's principle of geometrisation , since both the scalar and tensor fields have more or less intrinsic geometrical significance .In the consecutive investigations, Sen [7] and Sen and Dunn [7] proposed a new scalar tensor theory of gravitation and constructed an analog of the Einstein field equation based on Lyra's geometry which in normal gauge may be written as

$$R_{ik} - \frac{1}{2} g_{ik} R + (3/2) \phi_i \phi_k - \frac{3}{4} g_{ik} \phi_m \phi^m = - 8\pi T_{ik} \qquad \ldots(2)$$

where $\phi_i$ is the displacement vector and other symbols have their usual meaning as in Riemannian geometry.

Halford [8] has pointed out that the constant displacement field $\phi_i$ in Lyra's geometry play the role of cosmological constant $\Lambda$ in the normal general relativistic treatment. According to Halford the present theory predicts the same effects within observational limits, as far as the classical solar system tests are concerned, as well as tests based on the linearised form of field equations. Soleng [9] has pointed out that the constant displacement field in Lyra's geometry will either include a creation field and be equal to Hoyle's creation field cosmology or contain a special vacuum field which together with the gauge vector term may be considered as a cosmological term.

Subsequent investigations were done by several authors in scalar tensor theory and cosmology within the frame work of Lyra geometry [10].

Recently , I have studied some topological defects within the frame work of Lyra geometry[11] .

In this work we shall deal with cosmic strings with constant displacement vectors based on Lyra geometry in normal gauge i.e. displacement vector

$$\phi_i = ( \beta = \text{constant}, 0,0,0) \qquad \ldots\ldots(3)$$

and look forward whether the strings shows any significant properties due to introduction of the gauge field in the Riemannian geometry .

**Our paper is organized as follows :**

In section 2 , we consider a non static local string with energy momentum due to string $T_r^r = T_\theta^\theta = 0$ and $T_t^t = T_z^z \neq 0$ . In section 3 , we find an approximation solutions for the space time out side the core of the global string in Lyra geometry. In section 4 , the motion of the test particles are discussed . The paper ends with a short discussions in section 5 .

## 2. Local cosmic string in Lyra geometry :

The non static metric for the string is taken as [12]

$$ds^2 = e^{2A(r)}[dt^2 - e^{2b(t)}dz^2] - dr^2 - r^2 C^2(r) d\theta^2 \qquad \ldots\ldots(4)$$

The energy momentum due to string are taken to be [3]

$$T_r^r = T_\theta^\theta = 0 \text{ and } T_t^t = T_z^z = -\sigma \qquad \ldots\ldots\ldots(5)$$

$\sigma$ being the energy density of the string.

The field equation (1) for the metric (4) reduces to

$$A^{11} + (A^1)^2 + (C^{11}/C) + 2(C^1/rC) + A^1[(C^1/C) + (1/r)] + \tfrac{3}{4}\beta^2 e^{-2A} = 8\pi\sigma \qquad \ldots(6)$$

$$(b^{\bullet\bullet} + b^{\bullet 2})e^{-2A} - (A^1)^2 - 2A^1[(C^1/C) + (1/r)] - \tfrac{3}{4}\beta^2 e^{-2A} = 0 \qquad \ldots\ldots(7)$$

$$(b^{\bullet\bullet} + b^{\bullet 2})e^{-2A} - 3(A^1)^2 - 2A^{11} - \tfrac{3}{4}\beta^2 e^{-2A} = 0 \qquad \ldots\ldots(8)$$

[ dot and prime denote the differentiation w.r.t. 't' and 'r' respectively.]

From eq.(8), we get

$$[2A^{11} + 3(A^1)^2]e^{2A} + \tfrac{3}{4}\beta^2 = b^{\bullet\bullet} + b^{\bullet 2} = k \qquad \ldots\ldots(9)$$

( k is a separation constant)

For space part, we get

$$Z^{11} + \tfrac{1}{2}[(Z^1)^2/Z] = 2 A_0 \qquad \ldots\ldots\ldots(10)$$

Where $Z = e^{2A}$ and $A_0 = k - \tfrac{3}{4}\beta^2$.

Solving eq.(10), we get

$$Z = e^{2A} = \tfrac{1}{2} A_0 (r - r_0)^2 \qquad \ldots\ldots..(11)$$

($r_0$ is an integration constant)

From eqs.(7) & (8), one can get

$$C = [A^1 e^A / r] = [\sqrt{(\tfrac{1}{2} A_0)} / r] \qquad \ldots\ldots\ldots(12)$$

For time part, we get from eq.(9),

$$b(t) = \ln [\cosh\sqrt{k}\, t] \qquad \ldots\ldots\ldots(13)$$

Hence the space time of the local string is

$$ds^2 = \tfrac{1}{2} A_0 (r - r_0)^2 [dt^2 - \cosh^2\sqrt{k}\, t\, dz^2] - dr^2 - \tfrac{1}{2} A_0\, d\theta^2 \qquad \ldots\ldots(14)$$

Here the energy density $\sigma$ is of the form

$$8\pi\sigma = 3\beta^2 / [2A_0 (r - r_0)^2] \qquad \ldots\ldots(15)$$

We see that the solution of local string in Lyra geometry coincides with the asymptotic solution of the Gregory's non singular global string [12].

## 3. Global string in Lyra geometry :

For a global string, the energy momentum tensor components are calculated from the action density for a complex scalar field $\psi$ along with a Maxican hat potential :

$$L = \tfrac{1}{2} g^{ab} \psi^*_{,a} \psi_{,b} - (\lambda/4)(\psi^*\psi - v^2)^2 \qquad \ldots\ldots\ldots(16)$$

Where $\lambda, v$ are constants and $\delta = (v\sqrt{\lambda})^{-1}$ is a measure of the core radius of the string. It has been shown that the field configuration can be chosen as

$$\psi(r) = v f(r) \exp(i\theta) \qquad \ldots\ldots\ldots(17)$$

in cylindrical field co-ordinates.

The usual boundary conditions on $f(r)$ is $f(0) = 0$ and $f(r) \to 1$ as $r \to \delta$. As we are interested in space time outside the core of the string, for our purpose

$$f(r) = 1,\; f^1(r) = 0 \qquad \ldots\ldots\ldots(18)$$

is a good approximation.
The non zero components of the energy momentum tensor outside the core of the string now become [12]

$$T^t_t = T^r_r = T^z_z = -T^\theta_\theta = [v^2 / (2r^2 C^2)] \qquad \ldots\ldots\ldots(19)$$

The gravitational field equations (1) for a global string in Lyra geometry look like

$$A^{11} + (A^1)^2 + (C^{11}/C) + 2(C^1/rC) + A^1[(C^1/C) + (1/r)] + \tfrac{3}{4}\beta^2 e^{-2A} = -[4\pi v^2/(r^2 C^2)] \quad ..(20)$$

$$(b^{\bullet\bullet} + b^{\bullet 2}) e^{-2A} - (A^1)^2 - 2A^1[(C^1/C) + (1/r)] - \tfrac{3}{4}\beta^2 e^{-2A} = [4\pi v^2/(r^2 C^2)] \quad ...(21)$$

$$(b^{\bullet\bullet} + b^{\bullet 2}) e^{-2A} - 3(A^1)^2 - 2A^{11} - \tfrac{3}{4}\beta^2 e^{-2A} = -[4\pi v^2/(r^2 C^2)] \quad ..(22)$$

Clearly eqs.(21) & (22) imply

$$(b^{\bullet\bullet} + b^{\bullet 2}) - \tfrac{3}{4}\beta^2 = b_0 \quad ....(23)$$

[ $b_0$ is a constant ]

Hence

$$b(t) = \ln[\cosh\sqrt{(b_0 + \tfrac{3}{4}\beta^2)}\, t] \quad .....(24)$$

At this stage, let us consider the weak field approximations and assume

$$e^{2A} = 1 + f(r) \; ; \; C^2 = 1 + g(r) \quad .....(25)$$

Here the functions f, g should be computed to the first order in $v^2$ and $(b_0 + \tfrac{3}{4}\beta^2)$.

In these approximations, eqs.(20) – (22) take the following forms as

$$f^{11} + g^{11} + 2(g^1/r) + (f^1/r) + \tfrac{3}{4}\beta^2 = -4\pi v^2/r^2 \quad ...(26)$$

$$b_0 - (2f^1/r) = 4\pi v^2/r^2 \quad ...(27)$$

$$b_0 - 2f^{11} = -4\pi v^2/r^2 \quad ...(28)$$

Solving these equations, we get

$$f = \tfrac{1}{4} b_0 r^2 - 2\pi v^2 \ln r \quad ....(29)$$

$$g = -4\pi v^2 \ln r - (1/6)(b_0 + \tfrac{3}{4}\beta^2) r^2 \quad ......(30)$$

Thus in the weak field approximation, the non static global string in Lyra geometry takes the following form

$$ds^2 = (1 + \tfrac{1}{4} b_0 r^2 - 2\pi v^2 \ln r)[dt^2 - \cosh^2\sqrt{(b_0 + \tfrac{3}{4}\beta^2)}\, t\, dz^2] - dr^2 -$$

$$r^2 [1 - 4\pi v^2 \ln r - (1/6)(b_0 + \tfrac{3}{4}\beta^2) r^2] d\theta^2 \quad ....(31)$$

## 4. Gravitational effects on test particles:

Let us now consider a relativistic particle of mass m, moving in the gravitational field of global string described by equation (31) using the formalism of Hamilton and Jacobi (H – J).

According, the H – J equation is [13]

$$1/A(r)[\delta S/\delta t]^2 - 1/A(r)B(t)[\delta S/\delta z]^2 - [\delta S/\delta r]^2 - 1/C(r)[\delta S/\delta \varphi]^2 + m^2 = 0 \quad \ldots(32)$$

where $A = 1 + \frac{1}{4} b_0 r^2 - 2\pi v^2 \ln r$ ; $B = \cosh^2 \sqrt{(b_0 + \frac{3}{4}\beta^2)} \, t$

$$..(33)$$

$$C = r^2 [1 - 4\pi v^2 \ln r - (1/6)(b_0 + \frac{3}{4}\beta^2) r^2]$$

In order to solve the particle differential equation, let us use the separation of variables for the H – J function S as follows [13].

$$S(r, z, \theta, t) = S_1(t) + S_2(r) + J.\theta + M.z \quad \ldots\ldots(34)$$

Here the constant J is identified as the angular momentum of the particle.

The expressions for $S_1(t)$ & $S_2(r)$ are

$$S_1(t) = \in\!\int [E^2 + (M^2/B)]^{1/2} dt \quad \ldots(35)$$

$$S_2(r) = \in\!\int [(E^2/A) + m^2 - (J^2/C)]^{1/2} dr \quad \ldots(36)$$

Where E is the separation constant and can be identified as the energy of the particle [13].

Hence the trajectory of the particle [13]

$$z = \in\!\int (1/B)[E^2 + (M^2/B)]^{-1/2} dt \quad \ldots(37)$$

$$\theta = \in\!\int (1/C)[(E^2/A) + m^2 - (J^2/C)]^{-1/2} dr \quad \ldots(38)$$

$$\int [E^2 + (M^2/B)]^{-1/2} dt = \int (1/A)[(E^2/A) + m^2 - (J^2/C)]^{-1/2} dr \quad ..(39)$$

Thus the radial velocity of the particle

$$dr/dt = A[(E^2/A) + m^2 - (J^2/C)]^{1/2} / [E^2 + (M^2/B)]^{1/2} \quad \ldots(40)$$

The turning points of the trajectory are given by $dr/dt = 0$ and as a consequence the potential curves are

$$E/m = [(AJ^2/m^2C) - A]^{1/2} \qquad \ldots\ldots(41)$$

The extremals of the potential curve are given by

$$4r^2\{\pi v^2 m^2 + (1/6)(b_0 + \tfrac{3}{4}\beta^2)j^2\} + 8\pi v^2 m^2 \ln r = b_0 r^4 m^2 + 4j^2(1 - \pi v^2) \quad ..(42)$$

This equation has real solution as $1 - \pi v^2 > 0$.
So orbit of a test particle is bounded i.e. particles can be trapped by global string.
In other words global string exerts gravitational effects on test particle which is attractive in nature.

## 5. Concluding Remarks :

We see that the non static local string with energy momentum tensor components
$T_r^r = T_\theta^\theta = 0$ and $T_t^t = T_z^z \neq 0$ exists in Lyra geometry.
If we take $b = 0$, then our local non static metric becomes static. Thus local static string exists in Lyra geometry. But in Brans-Dicke theory of gravity local static string does not exist [14]. We also see that the solutions of our non static local string in Lyra geometry coincides with asymptotic solution of Gregory's non singular global string [12]. For non static global string in Lyra geometry, we have shown that global string exerts gravitational effects on test particle which is attractive in nature. For future study one can obtain non static string solution with time dependent displacement vector within the frame work of Lura geometry.

Acknowledgements :
                I am grateful to Prof.S.Chakraborty and Dr.A.A.Sen for helpful discussions. I am thankful to UGC for financial support and IUCAA for sending papers and preprints. I am also grateful to the referee for his valuable comments.